\documentclass[]{revtex4-2}
\usepackage{graphicx}
\usepackage{hyperref}
\usepackage[normalem]{ulem}
\usepackage{dcolumn}
\usepackage{multirow} 
\usepackage{xcolor}

\begin{document}
\baselineskip=5.0ex

\begin{center}
{\large\bf Size-consistency and orbital-invariance issues revealed by VQE-UCCSD calculations with the FMO scheme}
\par
{Kenji Sugisaki$^{a,b,c,\dagger}$
, Tatsuya Nakano$^{d}$, Yuji Mochizuki$^{e,f,\ddagger}$}
\par \smallskip
$^a${\it Graduate School of Science and Technology, Keio University, 7-1 Shinkawasaki, Saiwai-ku, Kawasaki, Kanagawa 212-0032, Japan} \par
$^b${\it Quantum Computing Center, Keio University, 3-14-1 Hiyoshi, Kohoku-ku Yokohama, Kanagawa 223-8522, Japan} \par
$^c${\it Centre for Quantum Engineering, Research and Education, TCG Centres for Research and Education in Science and Technology, Sector V, Salt Lake, Kolkata 700091, India} \par
$^d${\it Division of Medicinal Safety Science,
National Institute of Health Sciences, 1-18-1 Kamiyoga, Setagaya-ku, Tokyo 158-8501, Japan} \par
$^e${\it Department of Chemistry and Research Center for Smart Molecules,
Faculty of Science, Rikkyo University, 3-34-1 Nishi-ikebukuro, Toshima-ku, Tokyo 171-8501, Japan} \par
$^f${\it Institute of Industrial Science, The University of Tokyo, 4-6-1 Komaba,Meguro-ku, Tokyo 153-8505, Japan} \par
$^\dagger$Corresponding author: ksugisaki@keio.jp, $^\ddagger$fullmoon@rikkyo.ac.jp \par
\bigskip
{\bf 2nd version for arXiv, 2024/3/13, JST} 
\end{center}

\newpage
\noindent{\Large\bf Abstract} \par
The fragment molecular orbital (FMO) scheme is one of the popular fragmentation-based methods and has the potential advantage of making the circuit flat in quantum chemical calculations on quantum computers. In this study, we used a GPU-accelerated quantum simulator (cuQuantum) to perform the electron correlation part of the FMO calculation as unitary coupled-cluster singles and doubles (UCCSD) with the variational quantum eigensolver (VQE) for hydrogen-bonded (FH)$_3$ and (FH)$_2$-H$_2$O systems with the STO-3G basis set. VQE-UCCD calculations were performed using both canonical and localized MO sets, and the results were examined from the point of view of size-consistency and orbital-invariance affected by the Trotter error. It was found that the use of localized MO leads to better results, especially for (FH)$_2$-H$_2$O. The GPU acceleration was substantial for the simulations with larger numbers of qubits, and was about a factor of 6.7--7.7 for 18 qubit systems.
\par

\bigskip
\noindent{\Large\bf Keywords} \par
FMO, UCC, GPU, VQE, Trotter Error \par

\newpage
\noindent{\Large\bf 1. Introduction} \par
Starting with the seminal work of Aspuru-Guzik et al.~\cite{AAG-2005}, quantum chemical computation has been actively explored and developed as a promising application area for quantum computers~\cite{Cao-2019, McArdle-2020, Bauer-2020, Motta-2017, Bernal-2022}, where the potential applicability to huge-scale configuration interactions such as the FeMo-cofactor of nitrogenase~\cite{Reiher-2017} is attractive with care for the setting of active orbital space~\cite{Li-2019}.
In practice, however, the development of computational methods and algorithms using quantum simulators is currently more mainstream than the use of actual devices. For noisy intermediate-scale quantum (NISQ) computers, the unitary coupled-cluster singles and doubles (UCCSD)~\cite{Yaris-1964, Yaris-1965, Tanaka-1984, Bartlett-1989, Kutzelnigg-1991, Taube-2006, Harsha-2018, Anand-2022} has been used for relatively small molecules in conjunction with the variational quantum eigensolver (VQE)~\cite{Yung-2014, Peruzzo-2014, Romero-2018, Guo-2022}, and this VQE-UCCSD scheme has been extended to multi-reference cases, e.g., Refs.~\cite{Lee-2019, Stair-2020, Greene-Diniz-2020, Sugisaki-2022}. In addition, GPU-accelerated simulators e.g., cuQuantum~\cite{Bayraktar-2023} have attracted considerable interest due to its pronounced performance~\cite{Sugisaki-2023}. \par

In another direction, the so-called problem decomposition approach has been introduced to shallow the circuit depth~\cite{Dalton-2024} while avoiding the effects of noise. Note that such an approach is rather common for large molecules (like proteins) as the fragmentation-based methods~\cite{Gordon-2011, Collins-2015, Raghavachari-2015}. The introduction of problem decomposition to quantum computation was pioneered by Yamazaki et al.~\cite{Yamazaki-2018}, who compared three methods of fragment molecular orbital (FMO)~\cite{Kitaura-1999}, divide-and-conquer (DC)~\cite{Akama-2007}, and density matrix embedding theory (DMET)~\cite{Knizia-2012}.
Recently, the FMO calculations with VQE-UCCSD for hydrogen clusters have been reported~\cite{Lim-2024}. Note that the (H$_2$)$_2$, {\it trans}-butadiene, and [Cr$_2$(OH)$_3$(NH$_3$)$_6$]$^{3+}$ systems were treated by UCCSD based on the concept of orbital locality~\cite{Otten-2022}. \par

In this study, we applied the VQE-UCCSD scheme~\cite{Yung-2014, Peruzzo-2014, Romero-2018, Guo-2022} to compute the electron correlation part of FMO calculations~\cite{Kitaura-1999, Fedorov-2009, Mochizuki-2021} for a couple of hydrogen-bonded systems, (FH)$ {}_3 $ and (FH)$ {}_2 $-H$ {}_2 $O. For execution, the cuQuantum simulator~\cite{Bayraktar-2023} was used as done in the previous study~\cite{Sugisaki-2023}. Effect of Trotterization on the orbital-invariance condition of the UCCSD method was investigated using two symmetrically equivalent FH molecules in the latter system. We also studied relationship between the size-consistency condition and the Trotterized UCCSD ansatz, using square tetrahydrogen (4H) and cuboid octahydrogen (8H) clusters. Acceleration of the VQE-UCCSD simulations using cuQuantum is also discussed. The rest of the paper is organized as follows: In Section 2, we describe the calculation methods of both FMO stage and VQE-UCCSD stages. The results of the FMO correlation energies are shown first, and then the issues surrounding the Trotter error are discussed in Section 3. In Section 4, we summarize our work and discuss possible directions for future work. \par

\bigskip
\noindent{\Large\bf 2. Method of calculation} \par
\noindent{\bf 2.1. FMO scheme and program} \par
The scheme of the basic two-body FMO calculation~\cite{Kitaura-1999, Fedorov-2009, Mochizuki-2021} is summarized as follows. The first step is to determine the molecular orbital and electron density of each monomer by the Hartree--Fock (HF) approximation~\cite{Szabo-1982} under a given basis function, while self-consistently imposing an electrostatic potential (ESP) on each other. The set of ESPs of the monomers is to be determined until the monomer self-consistent-charge (SCC) condition is satisfied by iterations. This allows the polarization of each monomer to be taken into account. In the next step, the monomer-determined ESP is used to calculate the HF for the dimer; no SCC condition is imposed. The dimer calculation takes into account the delocalization of electrons between the monomers. From the sum of the HF energies of the monomer 
and the dimer, the two-body FMO energy of the system of interest is given as in Eq. (\ref{eq:eq1})
\begin{eqnarray}
E^{\rm FMO} = \sum_{I>J} E_{IJ} 
- (N_f -2) \sum_I E_I. 
\label{eq:eq1}
\end{eqnarray}
Indices of $ I $ and $J $ specify the respective monomers, and $ N_f $ is the number of fragments. \par

Electron correlation calculations, such as second-order M{\o}ller--Plesset perturbation (MP2)~\cite{Szabo-1982}, are performed after the HF calculations for each monomer are complete and after the individual HF calculations for each dimer are complete. The correlation energy correction is done in an additive manner as in Eq. (\ref{eq:eq1}).
The introduction of electron correlations is essential to improve quantitatively by incorporating dispersion stabilization and reducing excess ionicities of the HF description. As described in Ref.~\cite{Szabo-1982}, both size-consistency and orbital-invariance are crucial requirements in the correlated methods. This is obviously true for the FMO scheme based on Eq. (\ref{eq:eq1}). \par

Currently, GAMESS-US~\cite{Fedorov-2017, Fedorov-2021}, PAICS~\cite{Ishikawa-2009, Ishikawa-2021}, and ABINIT-MP~\cite{Tanaka-2014, Mochizuki-2021a} are the available programs that can perform FMO calculations including electron correlation correction by MP2. Besides the MP2 capability~\cite{Mochizuki-2004b, Mochizuki-2004a, Mochizuki-2008}, ABINIT-MP is unique in supporting higher-order correlated calculations~\cite{Szabo-1982, Shavitt-2009} on-the-fly; from the third-order MP (MP3)~\cite{Mochizuki-2010} 
to coupled-cluster singles and doubles including perturbative triples (CCSD(T))~\cite{Mochizuki-2011} are supported. \par

\medskip
\noindent{\bf 2.2. Preparation of molecular integrals under FMO scheme} \par
\begin{figure}
\includegraphics[width=70mm]{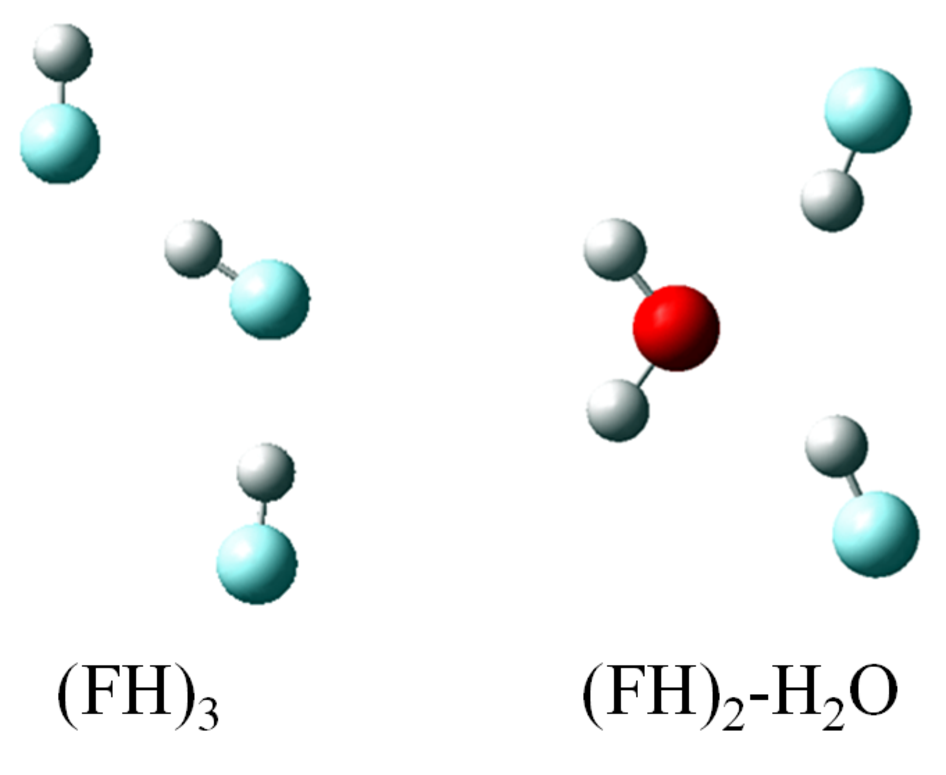}
\caption{\label{fig:fig1} Molecular structures of (FH)$_3$ and (FH)$_2$-H$ _2$O. For the former, the middle, upper, and lower FH molecules correspond to fragments ``1'', ``2'', and ``3'', respectively. For the latter, the H$_2$O molecule is assigned to fragment ``1''; two FH molecules (fragments ``2'' (upper) and ``3'' (lower)) are equivalent due to the C$_{\rm 2v}$ symmetry.}
\end{figure}

The geometries of (FH)$ {}_3 $ (under C$ {}_{\rm s } $ symmetry) and (FH)$ {}_2 $-H$ {}_2 $O (C$ {}_{\rm 2v} $ symmetry) were optimized by the GAUSSIAN16W program~\cite{GAUSSIAN-2016} at the level of B3LYP~\cite{Becke-1993} corrected with the empirical dispersion~\cite{Grimme-2010} with the 6-31+G(d',p') basis set~\cite{Petersson-1991}. The resulting Cartesian coordinates are listed in TABLE~\ref{tab:table1} and illustrated in FIG.~\ref{fig:fig1}. \par

\begin{table}
\caption{\label{tab:table1} Optimized Cartesian coordinates in units of {\AA}. 
}
\begin{center}
\begin{tabular*}{10cm}{@{\extracolsep{\fill}}cccrrr}
\hline
Seq. & Frag. & Elem. & \multicolumn{1}{c}{x} & \multicolumn{1}{c}{y} & \multicolumn{1}{c}{z} \\
\hline
(FH)$ {}_3 $ &   &     &            &             \\
1 & 1 & F &   0.779023 &   0.287467 &   0.000000  \\
2 & 1 & H &   0.000000 &   0.807300 &   0.000000  \\
3 & 2 & F &  $-$1.361653 &   1.874668 &   0.000000  \\
4 & 2 & H &  $-$1.330503 &   2.801407 &   0.000000  \\
5 & 3 & F &   0.648089 &  $-$2.399606 &   0.000000  \\
6 & 3 & H &   0.741364 &  $-$1.471463 &   0.000000  \\
(FH)$ {}_2 $-H$ {}_2 $O &  &  &     &             \\
1 & 1 & O &   0.000000 &   0.000000 &   1.200039  \\
2 & 1 & H &   0.000000 &   0.771251 &   1.779594  \\
3 & 1 & H &   0.000000 &  $-$0.771251 &   1.779594  \\
4 & 2 & F &   0.000000 &   2.034006 &  $-$0.694256  \\
5 & 2 & H &   0.000000 &   1.179139 &  $-$0.331451  \\
6 & 3 & F &   0.000000 &  $-$2.034006 &  $-$0.694256  \\
7 & 3 & H &   0.000000 &  $-$1.179139 &  $-$0.331451  \\
\hline 
\end{tabular*}
\end{center}
\end{table}

For (FH)$ {}_3 $ and (FH)$ {}_2 $-H$ {}_2 $O, the FMO calculations were performed with the STO-3G minimal basis set~\cite{Hehre-1969}, where we used a local version of ABINIT-MP, which dumped the integral list of basis functions and the converged canonical MO (CMO) coefficients (at the FMO-HF level) of monomers and dimers as separate files.
These data were transformed by a small Fortran program into molecular integrals for the second-quantized Hamiltonian used to run VQE-UCCSD, expressed as
\begin{eqnarray}
H = \sum_{pq} h_{pq} a_p^{\dagger} a_q + 
\frac{1}{2} \sum_{pqrs} g_{pqrs} a_p^{\dagger} a_q^{\dagger} a_s a_r . 
\label{eq:eq2}
\end{eqnarray}
Indices of $ p,~q,~r $ and $ s $ cover the correlating orbital space, and $ h_{pq} $ and $ g_{pqrs} $ are the transformed one- and two-electron integrals.
The 1s-like CMOs of fluorine and oxygen were frozen~\cite{Hosteny-1975} for $ h_{pq} $ in Eq. (\ref{eq:eq2}), which is a good approximation to save on the number of 
qubits~\cite{McCaskey-2019, Mochizuki-2019}. The number of correlated electrons for dimers was thus 16.\par

Note that there is some degree of locality of monomer CMOs in dimer orbitals with respect to the occupied space for (FH)$ {}_3 $ and also that there is the symmetric delocalization for the FH dimer in (FH)$ {}_2 $-H$ {}_2 $O. To address the issue of size-consistency, the Pipek--Mezey localization\cite{PML-1989} was performed for the valence occupied CMOs and the virtual CMOs, respectively, and these sets of localized MOs (LMOs) were also used for the integral transformation. The lists of molecular orbitals (CMOs and LMOs) of for the monomers and dimers of (FH)$_3$ and (FH)$_2$-H$_2$O are shown in FIG. S1 and S2, respectively, in Supplementary Materials. \par

Due to a proof-of-concept (PoC) phase of this study, the FMO calculations (at the HF level) by ABINIT-MP were done in a separate step from the quantum calculations described in the next subsection. For comparison with the VQE-UCCSD correlation energies, the usual FMO-MP2 and FMO-CCSD(T) calculations were also performed by ABINIT-MP. These calculations were completed in less than 1 second on a single core of Intel Xeon processor. \par

\medskip
\noindent{\bf 2.3. Set-up of VQE-UCCSD calculation} \par
VQE is a quantum--classical hybrid algorithm and it has been proposed to solve quantum chemistry problems using NISQ devices~\cite{Yung-2014, Peruzzo-2014}. In VQE, an approximate wave function is generated by using a parameterized quantum circuit (PQC) defined by an ``ansatz'', and the expectation value of the qubit Hamiltonian obtained by applying the fermion--qubit transformation to the second-quantized Hamiltonian given in Eq. (\ref{eq:eq2}) is computed statistically, by repeatedly executing the quantum circuit and collecting the measurement results. The classical computer then execute a variational optimization of the parameters in PQC. These steps are iterated until convergence. \par
Various types of ansatzes have been proposed and studied for quantum chemical calculations~\cite{Tilly-2022}. In this work, we adopted the UCCSD ansatz defined in Eqs. (\ref{eq:eq3}) and (\ref{eq:eq4}), because it is a chemically motivated ansatz and it can give very accurate correlation energies.
\begin{eqnarray}
|\Psi_{\rm{UCCSD}}\rangle = e^{T - T^\dagger}|\Psi_{\rm{HF}}\rangle. 
\label{eq:eq3}
\end{eqnarray}
\begin{eqnarray}
T = \sum_{ia} t_i^a a_a^\dagger a_i + \frac{1}{2}\sum_{ijab} t_{ij}^{ab} a_a^\dagger a_b^\dagger a_j a_i. 
\label{eq:eq4}
\end{eqnarray}
Here, we used the indices $i$ and $j$ for the occupied spin orbitals and $a$ and $b$ for the unoccupied orbitals of the HF wave function $|\Psi_{\rm{HF}}\rangle$. 
To accelerate the VQE simulations, we adopted the following techniques: (1) Using the symmetry conserving Bravyi--Kitaev transformation (SCBKT)~\cite{Bravyi-2017} to reduce two qubits in the simulation, (2) using the MP3 and the MP2 excitation amplitudes as the initial guess of the $t_i^a$ and $t_{ij}^{ab}$, respectively~\cite{Sugisaki-2022}, and (3) GPU-based numerical simulations. The number of qubits for the VQE-UCCSD simulations was 8 and 18 for monomers and dimers, respectively, in (FH)$_3$, and 10 for monomer ``1'' and 20 for dimers ``21'' and ``31'' in (FH)$_2$-H$_2$O. The VQE-UCCSD simulation program was developed by us, by using Python3 with OpenFermion~\cite{McClean-2020}, Cirq~\cite{Cirq}, and cuQuantum~\cite{Bayraktar-2023} libraries. \par

In the implementation of the UCCSD quantum circuit, we adopted the first-order Trotter decomposition given in Eq. (\ref{eq:eq5}) in conjunction with the magnitude ordering~\cite{Tranter-2018} of the cluster operators. 
\begin{eqnarray}
\mathrm{exp}\left(T - T^\dagger \right) = 
\mathrm{exp}\left(\sum_{k=1}^K i t_k P_k\right) \approx \left[ \prod_{k=1}^K e^{i t_k P_k/M} \right]^M
\label{eq:eq5}
\end{eqnarray}
Here, $\sum_{k=1}^K i t_k P_k$ is the excitation/de-excitation operators in the Pauli operator expressions obtained by adopting the SCBKT to the operator $(T - T^{\dagger})$ in the second-quantized form. $P_k$ is a direct product of Pauli operators called as a Pauli string, and $t_k$ is the corresponding coefficient derived from $t_i^a$ and $t_{ij}^{ab}$. \textcolor{red}{$K$} is the number of Pauli strings, and $M$ is the number of Trotter slices. Unless otherwise specified we used the one Trotter slice ($M = 1$) for the VQE-UCCSD simulations. 

For the variational optimization of the excitation amplitudes, we examined COBYLA~\cite{COBYLA} and Powell~\cite{Powell} algorithms; the corresponding labels are shortly denoted as UCCSD:CB and UCCSD:PW, respectively (see TABLE \ref{tab:table2} and \ref{tab:table3}). For comparison, the calculation
of the complete active space configuration interaction (CAS-CI) was performed to obtain the exact correlation energy in the orbital space of STO-3G.
The numerical simulations for (FH)$_3$ and (FH)$_2$-H$_2$O were carried out on the Supercomputer `Flow' Type-II subsystem at Nagoya University and on the in-house NVIDIA DGX H100 system, respectively.\par

\bigskip
\noindent{\Large\bf 3. Results and discussion} \par
\noindent{\bf 3.1. Energies and timings} \par
The correlation energies for (FH)$ {}_3 $
are summarized in TABLE~\ref{tab:table2}. Compared to MP2, CCSD has a significantly lower energy, and CCSD(T) gives values close to CAS-CI, as expected. UCCSD:PW gave lower energies than UCCSD:CB, but the number of function evaluations (total energy calculations) required in Powell is about 1.6--2.8 times greater than in COBYLA (see TABLE S1 in Supplementary Materials for details). The same trend was observed for the LMO-based UCCSD calculations. No significant difference was found in the number of function evaluations between CMO and LMO-based calculations.

As we discuss in the next section, Trotterized UCCSD does not automatically satisfy the size-consistency condition, and using LMOs as the basis is crucial to ensure that Trottterized UCCSD is size-consistent. In fact, the correlation energies of the dimers are improved in the LMO-based UCCSD:PW calculations, and the sum of the correlation energies is 0.001679 Hartree (1.0536 kcal mol$^{-1}$) lower in the LMO-based calculations than in the CMO-based one. The deviation of the sum of UCCSD:PW correlation energy from the CAS-CI one is 0.71 kcal mol$^{-1}$.

\begin{table}
\caption{\label{tab:table2} Correlation energies of (FH)$_3$\footnote{The HF energies (in units of Hartree) of monomer ``1'', ``2'', and ``3'' are $-$103.815720, $-$103.995064, $-$103.563842, respectively. In contrast, the HF energies of dimer ``21'', ``31'', ``32'' are $-$228.173251, $-$227.542281, $-$218.792929, respectively. The sum of Eq. (1) is $-$363.133835 Hartree.
Eq. (1) is also used for the sum of correlation energies.} in units of Hartree. 
}
\begin{center}
\begin{tabular*}{16cm}{@{\extracolsep{\fill}}ccccccccc}
\hline
Unit&    MP2&      CCSD&     CCSD(T)&  \multicolumn{2}{c}{UCCSD:CB} & \multicolumn{2}{c}{UCCSD:PW} & CAS-CI \\
&&&& CMO\footnote{HF canonical orbitals were used for the calculation. } & LMO\footnote{Localied molecular orbitals constructed by using Pipek--Mezey method were used for the calculation.} & CMO & LMO & \\
\hline  
Monomer	&&&&&&&& \\  
``1''  &  $-$0.017933 & $-$0.026945 & $-$0.026945 & $-$0.026884 & $-$0.026854 & $-$0.026914 & $-$0.026729 & $-$0.026945\\
``2''  &  $-$0.017526 & $-$0.026216 & $-$0.026216 & $-$0.026164 & $-$0.026169 & $-$0.026192 & $-$0.026019 & $-$0.026216\\
``3''  &  $-$0.017933 & $-$0.026929 & $-$0.026929 & $-$0.026899 & $-$0.026839 & $-$0.026875 & $-$0.026691 & $-$0.026929\\
Dimer &&&&&&&& \\		                  	                  	     	
``21'' &  $-$0.035493 & $-$0.051856 & $-$0.051933 & $-$0.049880 & $-$0.050429 & $-$0.050452 & $-$0.051322 & $-$0.051963\\
``31'' &  $-$0.035980 & $-$0.052778 & $-$0.052850 & $-$0.051554 & $-$0.051632 & $-$0.051527 & $-$0.052169 & $-$0.052879\\
``32'' &  $-$0.035446 & $-$0.053122 & $-$0.053123 & $-$0.051952 & $-$0.051317 & $-$0.053067 & $-$0.052692 & $-$0.053124\\
 &&&&&&&& \\
Sum.   &  $-$0.053527 & $-$0.077666 & $-$0.077816 & $-$0.073439 & $-$0.073516 & $-$0.075065 & $-$0.076744 & $-$0.077876 \\
\hline
\end{tabular*}
\end{center}
\end{table}

TABLE~\ref{tab:table3} summarizes the results for the (FH)$ {}_2 $-H$ {}_2 $O correlation energies. The number of function evaluations in the VQE-UCCSD optimization is given in TABLE S2 in Supplementary Materials.
A checkpoint here is whether the equivalence symmetries (monomers ``2'' and ``3'' / dimers ``21'' and ``31'') are satisfied, and the usual MP2, CCSD, CCSD(T), and CAS-CI results all satisfy this requirement. The trend of the correlation energies by these methods is the same as in (FH)$ {}_3 $. On the other hand, the UCCSD results (of both COBYLA and Powell) unfortunately do not satisfy symmetry, as the difference is seen to five decimal places for monomers and three decimal places (in the order of kcal mol$^{-1}$) for dimers. From a chemical precision point of view, it seems problematic that the effect is seen to three decimal places.
Furthermore, this issue of broken equivalence should be kept in mind not only for FMO, but for all approaches of fragmentation-oriented methods~\cite{Gordon-2011, Collins-2015, Raghavachari-2015}.
This problem of VQE-UCCSD is related to the Trotter error and is discussed in the next section. 
 
The effect of orbital localization on the UCCSD correlation energy is remarkable in the (FH)$_2$-H$_2$O system. In the UCCSD:PW calculations the sum of correlation energies improved about 0.005256 Hartree (3.2982 kcal mol$^{-1}$) by the orbital localization, and deviation from the CAS-CI correlation energy is 1.29 kcal mol$^{-1}$. Note that orbital localization also improves the orbital-invariance condition. By using the LMOs, the difference in correlation energy between ``21'' and ``31'' is reduced to 0.039 kcal mol$^{-1}$. These results exemplify the importance of using LMOs in the combination of FMO and VQE-UCCSD approaches. 

\begin{table}
\caption{\label{tab:table3} Correlation energies of (FH)$ {}_2 $-H$ {}_2 $O\footnote{The HF energies in units of Hartree of monomer ``1'', ``2'', and ``3'' are $-$84.426515, $-$103.695254, and $-$103.695254, respectively (``2'' and ``3'' are equivalent). 
In contrast, the HF energies of dimer ``21'', ``31'', ``32'' are $-$207.357701, $-$207.357701, and $-$220.935631, respectively (``21'' and ``31'' are equivalent). The sum of Eq. (1) is $-$343.834010 Hartree. Eq. (1) is also used for the sum of correlation energies.} in units of Hartree.
}
\begin{center}
\begin{tabular*}{16cm}{@{\extracolsep{\fill}}ccccccccc}
\hline
Unit&    MP2&      CCSD&     CCSD(T)&  \multicolumn{2}{c}{UCCSD:CB}& \multicolumn{2}{c}{UCCSD:PW}& CAS-CI \\
&&&& CMO\footnote{HF canonical orbitals were used for the calculation. } & LMO\footnote{Localied molecular orbitals constructed by using Pipek--Mezey method were used for the calculation.} & CMO & LMO & \\
\hline
Monomer	&&&&&&&& \\                          	                  	
``1''  & $-$0.035370 & $-$0.049321 & $-$0.049394 & $-$0.049242 & $-$0.049214 & $-$0.049231 & $-$0.048915 & $-$0.049445\\
``2''  & $-$0.017810 & $-$0.026705 & $-$0.026705 & $-$0.026608 & $-$0.026648 & $-$0.026692 & $-$0.026538 & $-$0.026705\\
``3''  & $-$0.017810 & $-$0.026705 & $-$0.026705 & $-$0.026677 & $-$0.026637 & $-$0.026659 & $-$0.026476 & $-$0.026705\\
Dimer &&&&&& \\		                  	                  	     	
``21'' & $-$0.053486 & $-$0.075392 & $-$0.075526 & $-$0.069313 & $-$0.071305 & $-$0.071421 & $-$0.074429 & $-$0.075600\\
``31'' & $-$0.053486 & $-$0.075392 & $-$0.075526 & $-$0.071847 & $-$0.069583 & $-$0.072544 & $-$0.074367 & $-$0.075600\\
``32'' & $-$0.035790 & $-$0.053549 & $-$0.053553 & $-$0.053230 & $-$0.052161 & $-$0.053207 & $-$0.052979 & $-$0.053549\\
 &&&&&&&& \\
Sum.   & $-$0.071770 & $-$0.101602 & $-$0.101801 & $-$0.091863 & $-$0.090551 & $-$0.094590 & $-$0.099846 & $-$0.101895\\
\hline
\end{tabular*}
\end{center}
\end{table}

The cuQuantum quantum simulator was used in this VQE-UCCSD computation. 
TABLE~\ref{tab:table4} summarizes the timings of the UCCSD jobs of (FH)$_3$ using LMOs on the `Flow' Type-II subsystem with and without GPU. The GPU acceleration was about a factor of 1.6--2.2 and 6.7--7.7 for monomers and dimers, respectively. The VQE-UCCSD simulations of the dimers ``21'' and ``31'' of (FH)$_2$-H$_2$O (20 qubit systems) without GPU acceleration are too time-consuming to do. Here we estimated the acceleration ratio by performing the UCCSD simulations of the dimer ``21'' of (FH)$_2$-H$_2$O by setting the maximum number of function evaluations in the VQE parameter optimization to be 100 on NVIDIA DGX H100. With the GPU acceleration, the time required for pre-processing (Fermion--qubit transformation, reference CAS-CI calculation, MP2 and MP3 calculations, etc.) was 4773.9 seconds and 100 function evaluations in the VQE optimization took 1211.0 seconds. In contrast, the CPU-only calculation took 4421.2 and 37363.9 seconds for pre-processing and 100 function calls, respectively. 
The GPU acceleration of the VQE iteration part is about a factor of 30.85. 
Since the numbers of function evaluations required for convergence in UCCSD:CB and UCCSD:PW were 4913 and 12452, respectively, the time for the CPU-only simulations are estimated to be about 21 and 54 days for UCCSD:CB and UCCSD:PW, respectively. GPU acceleration is substantial for larger systems, but the speedup is less significant compared with our previous study~\cite{Sugisaki-2023}. This is because the VQE job needs a lot of time for pre-processing and post-processing, and these parts cannot be accelerated by cuQuantum. Considering that a normal FMO-CCSD(T)/STO-3G calculation takes less than 1 second to complete, there is a speed difference of the order of the fourth power of 10 if the correlation part is due to VQE-UCCSD at this time. Note that the present VQE-UCCSD was run on a classical computer, where computational time grows exponentially with the number of qubits. 
Anyway, as in the previous report~\cite{Sugisaki-2023}, 
GPU acceleration with cuQuantum is essential for quantum simulations.

\begin{table}
\caption{\label{tab:table4} Timings of UCCSD job (in second) of (FH)$_3$ using LMOs with/without GPU.\footnote{All the calculations were carried out on `Flow' Type-II subsystem.} }
\begin{center}
\begin{tabular*}{15cm}{@{\extracolsep{\fill}}ccccccc}
\hline
Unit  & \multicolumn{3}{c}{UCCSD:CB} & \multicolumn{3}{c}{UCCSD:PW} \\
& with GPU & without GPU & Acceleration & with GPU & without GPU & Acceleration \\
\hline
Monomer &&&&&& \\
``1'' & 17.2 & 27.8 & 1.62 & 27.6 & 48.3 & 1.75 \\
``2'' & 15.5 & 33.6 & 2.17 & 28.5 & 55.4 & 1.94 \\
``3'' & 16.2 & 26.8 & 1.65 & 26.5 & 49.3 & 1.86 \\
Dimer &&&&&& \\
``21'' & 12767.5 & 93205.0 & 7.30 & 34153.3 & 260996.7 & 7.64 \\
``31'' & 13947.1 & 94199.7 & 6.75 & 34397.3 & 237695.3 & 6.91 \\
``32'' &  8805.9 & 59975.6 & 6.81 & 21129.0 & 156257.2 & 7.40 \\
\hline 
\end{tabular*}
\end{center}
\end{table}

\medskip
\noindent{\bf 3.2. Relation with Trotter error} \par

\begin{figure}
\includegraphics[width=10cm]{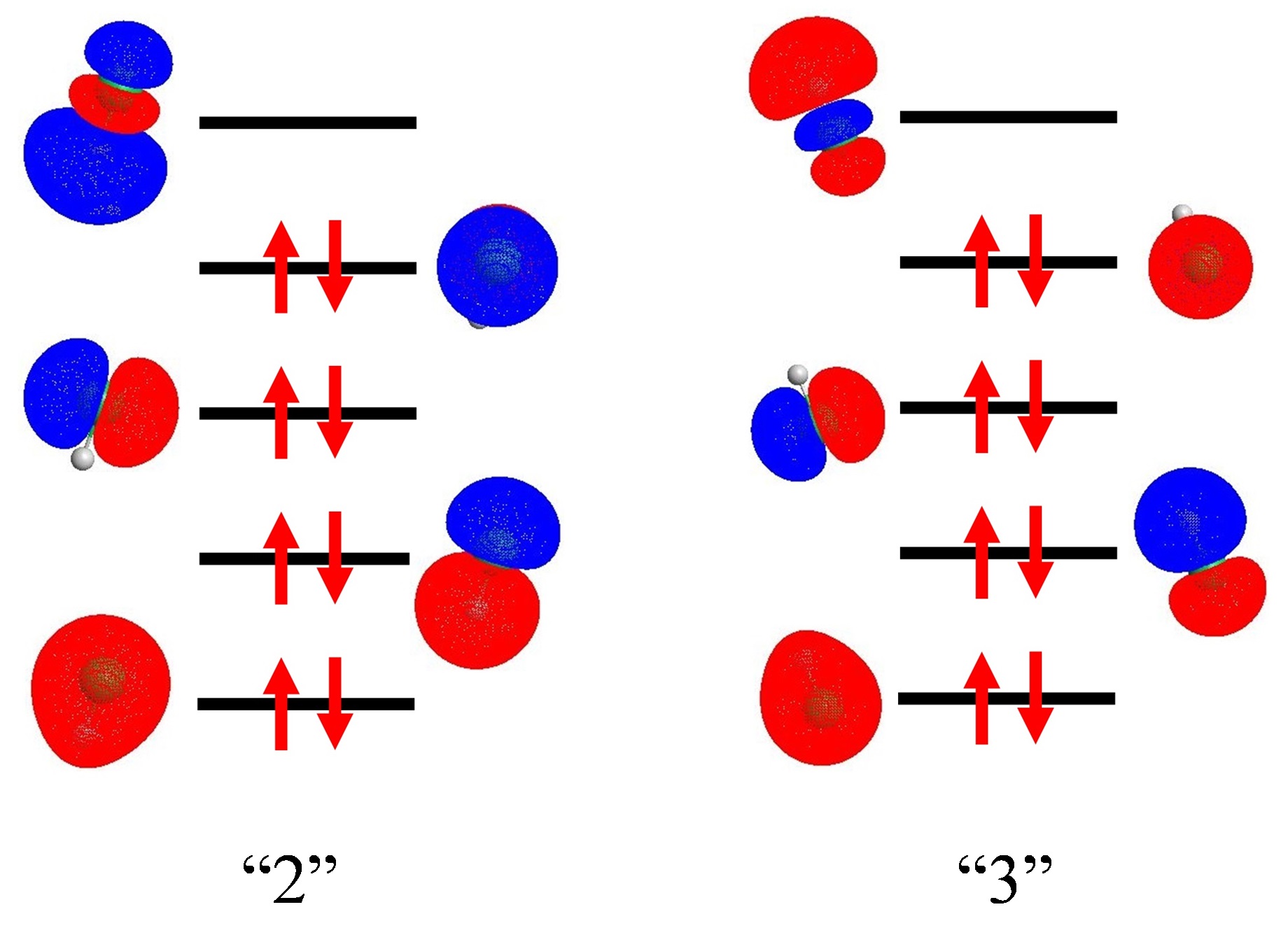}
\caption{\label{fig:fig2} Active orbitals of FH molecules (monomers ``2'' and ``3'') of (FH)$_2$-H$_2$O. Red arrows specify the electron occupancy of the HF wave function.}
\end{figure}

It is interesting to note that the monomers ``2'' and ``3'' of (FH)$_2$-H$_2$O are symmetrically equivalent, but VQE-UCCSD yields different correlation energies. The HF canonical orbitals of the monomers ``2'' and ``3'' are illustrated in FIG.~\ref{fig:fig2}. We found that the relative phase from the second to the fifth molecular orbitals is different (or inverted) between monomers ``2'' and ``3'', which causes changes in the absolute sign of the some excitation amplitudes $t_i^a$ and $t_{ij}^{ab}$. As a result, the quantum states corresponding to the UCCSD wave function are not identical for monomers ``2'' and ``3'', and the Trotter error appears in a different way. This fact is confirmed by performing the UCCSD calculations without Trotterization, using the \texttt{expm\_multiply} function in the SciPy library~\cite{SciPy}, which allows us to compute the action of the matrix exponential of $(T - T^\dagger)$ on $|\Psi_{\rm HF}\rangle$. In this case, the calculated correlation energies of the monomers ``2'' and ``3'' are exactly the same: $-0.026687$ Hartree. The fact that Trotterized UCCSD can not maintain orbital-invariance indicates that care must be taken to ensure that the relative phases of the molecular orbitals match at all points when calculating potential energy surfaces. 

Since the Trotter errors appears in an unexpected way, we further investigated about the relationship between Trotter errors and the size-consistency, which is an essential condition in the FMO framework as mentioned earlier. 
Here we focused on the tetrahydrogen (4H) cluster~\cite{Paldus-1993} in a square coordinate with {\it R}(H--H) = 1.0583 {\AA} (2.0 Bohr) as the monomer, because the Trotter error becomes more significant when the HF is not a good approximation of the ground-state wave function and the UCCSD wave function has large excitation amplitudes. This system is also suitable because the HF CMOs are completely defined by point-group symmetry. In the dimer (8H) calculations, two 4H clusters were placed to form a cuboid, with the inter-monomer distance being 100 {\AA}. Two types of molecular orbitals are examined in the dimer calculations: Completely delocalized canonical orbitals by HF in D$_{\rm 2h}$ point group and LMOs on the monomers. In the total energy calculations using UCCSD:PW, we used cuQuantum-based quantum circuit simulations with Trotter decomposition and without Trotter decomposition using the \texttt{expm\_multiply} function in SciPy. The results are summarized in TABLE~\ref{tab:table6}. \par

\begin{table}
\caption{\label{tab:table6} Deviations of the UCCSD:PW total energy from the CAS-CI value for 4H cluster (monomer) and 8H cluster (dimer).}
\begin{center}
\begin{tabular*}{10cm}{@{\extracolsep{\fill}}ccc}
\hline
Unit    & Trotter decomposition & $\Delta E/\rm{kcal\ mol^{-1}}$ \\
\hline
Monomer & No & 0.8118 \\
Dimer (LMO) & No & 1.6236 \\
Dimer (CMO) & No & 1.6234 \\
Monomer & Yes & 0.8102 \\
Dimer (LMO) & Yes & 1.6207 \\
Dimer (CMO) & Yes & 5.0319 \\
\hline 
\end{tabular*}
\end{center}
\end{table}

From TABLE~\ref{tab:table6}, the UCCSD calculations without Trotter decomposition yield almost the same $\Delta E$ values for both LMO- and CMO-based calculations, and the $\Delta E$ values of the dimer are twice those of the monomer; i.e., Trotter-free UCCSD satisfies the size-consistency condition. Small differences in the $\Delta E$ values of the dimer with CMO and LMO are due to rounding errors in the AO $\rightarrow$ MO transformation. In contrast, when the Trotter decomposition is used to construct the UCCSD quantum circuit, the $\Delta E$ of the dimer with CMO is significantly larger than the $2\times \Delta E\rm{(Monomer)}$. Since the $\Delta E$ value for dimer with LMO is approximately twice the $\Delta E$ of monomer, we concluded that the size-consistency condition of the VQE-UCCSD can be maintained when the molecular orbitals localized on each monomer are used in the calculations. 

In the present study, we used the first-order Trotter decomposition given in Eq. (\ref{eq:eq5}), with the number of Trotter slices $M = 1$. 
To further investigate the relationship between Trotter error and the size-consistency, we run the UCCSD:PW simulations with $M$ changed from 1 to 5. We also carried out the UCCSD:PW simulations with the second-order Trotter decomposition given in Eq. (\ref{eq:eq6}). 
\begin{eqnarray}
\mathrm{exp}\left(T - T^\dagger \right) = 
\mathrm{exp}\left(\sum_{j=1}^J i t_j P_j\right) \approx \left[ \prod_{j=1}^J e^{i t_j P_j/2M} \prod_{j=J}^1 e^{i t_j P_j/2M} \right]^M
\label{eq:eq6}
\end{eqnarray}
The results are summarized in TABLE S3 in the Supplementary Materials. The simulations ended within one hour when GPU is used. The UCCSD:PW energies of monomer (4H cluster) and dimer (8H cluster) with LMO do not change by using a larger number of Trotter slices or by adopting the second-order Trotter decomposition. For the dimer calculations with CMO, in contrast, the deviation from the CAS-CI energy systematically decreases with increasing $M$. However, even for $M = 5$, the UCCSD:PW energy does not converge to the Trotter-free UCCSD energy calculated by using \texttt{expm\_multiply} in SciPy. This result also implies the importance of using LMO in the FMO scheme in conjunction with the VQE-UCCSD. 

\begin{figure}
\includegraphics{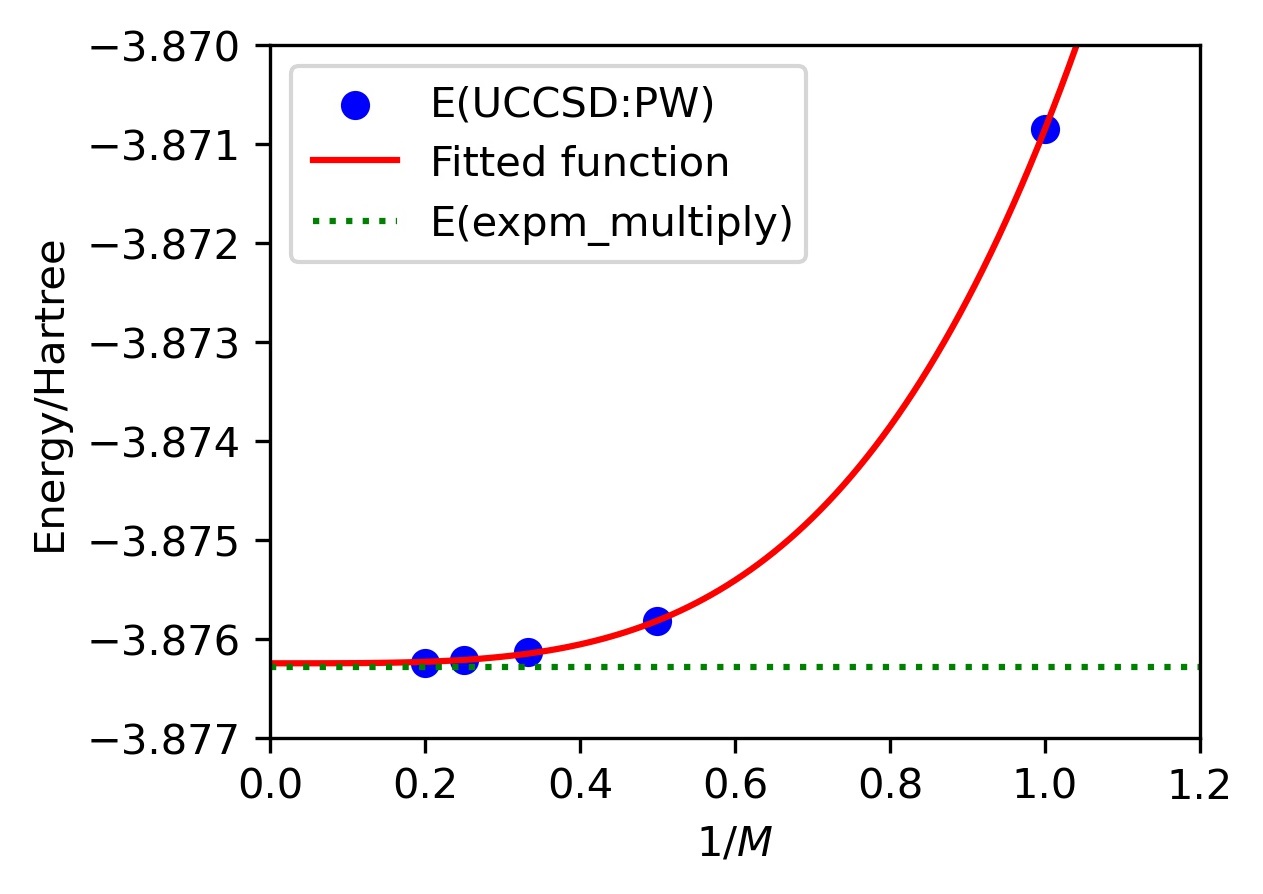}
\caption{\label{fig:fig3} The plot of the UCCSD:PW energies of 8H cluster with different number of Trotter slices $M$ and the result of extrapolation. }
\end{figure}

Since size-consistency is pivotal not only for FMO but also for general quantum chemical calculations, it is important to provide methods to estimate the Trotter error-free energy. Here we examined the extrapolation method to infer the Trotter error-free UCCSD energy using the idea of algorithmic error mitigation~\cite{Endo-2019}. To do this, we plotted the UCCSD:PW energies as a function of the inverse of the number of Trotter slices, $1/M$, and fitted with a function $E = \alpha(1/M)^\beta + \gamma$. It should be noted that in the context of Hamiltonian simulations, the error of the first-order Trotter decomposition scales as $O(1/M)$~\cite{Nielsen-Chuang}. In VQE, however, different Trotterized versions of the UCCSD correspond to different ansatzes, and thus the optimal variational parameters are different. Therefore, it is not necessary to scale the Trotter error of UCCSD as $O(1/M)$. The calculation results are illustrated in FIG.~\ref{fig:fig3}. The E(UCCSD:PW) were successfully fitted by the function $E = 0.005396(1/M)^{3.6466} - 3.876244$, and the difference between the energies estimated from the extrapolation and the Trotter error-free one calculated with \texttt{expm\_multiply} is only 35 $\mu$Hartree. We expect that the extrapolation method used in this work will help to obtain the VQE-UCCSD energy with the size-consistency condition. 

It should be noted that the dependence of the Trotter error on the locality of the molecular orbitals was investigated by Babbush and coworkers~\cite{Babbush-2015}, and they reported that the Trotter error is larger for localized orbital basis than for canonical orbitals and natural orbitals. The reported study focused only on single molecule (monomer), and our discussions are based on comparing the energies of monomers and a dimer. Our results do not contradict this previous study. 

\bigskip
\noindent{\Large\bf 4. Summary} \par
We have performed the VQE-UCCSD calculations~\cite{Anand-2022, Yung-2014, Peruzzo-2014, Romero-2018, Guo-2022} using the cuQuantum simulator~\cite{Bayraktar-2023} in conjunction with the FMO calculations for the (FH)$ {}_3 $ 
and (FH)$ {}_2 $-H$ {}_2 $O systems. 
The STO-3G minimal basis set~\cite{Hehre-1969} was used and the frozen-core restriction~\cite{Hosteny-1975} was imposed. By combining with symmetry conserving Bravyi--Kitaev transformation~\cite{Bravyi-2017}, we can simulate the HF$\cdots$H$_2$O dimer with 20 qubits. Both COBYLA and Powell methods were used for VQE optimization, 
with the latter usually providing better energies but requires more function calls. When the HF CMOs were used for the UCCSD wave function expansion, the calculated correlation energies were somewhat small in magnitude, possibly due to the breakdown of the size-consistency condition. By using the LMOs, the UCCSD correlation energies improved dramatically, and the differences in the correlation energies between the CAS-CI and the UCCSD in conjunction with the Powell optimizer were calculated to be 0.71 and 1.29 kcal mol$^{-1}$ for (FH)$_3$ and (FH)$_2$-H$_2$O, respectively. The (FH)$_2$-H$_2$O system has two symmetrically equivalent FH molecules that should have the same energies, but the Trotterized UCCSD does not satisfy symmetry and yields different energies. The difference in correlation energies between the dimers ``21'' and ``31'' is on the order of 1 kcal mol$^{-1}$ in the canonical orbital basis but it reduced to 0.039 kcal mol$^{-1}$ in the localized orbital basis. Size-consistency of the Trotterized UCCSD is also studied numerically using 4H and 8H model clusters. We found that the size-consistency condition can be broken when the molecular orbitals delocalized to the dimer are used for the calculation, and using molecular orbitals localized to the monomers is essential to satisfy the size-consistency. These findings on the relationship between size-consistency and orbital-invariance~\cite{Szabo-1982, Shavitt-2009} and the error in the Trotter decomposition are very important not only for FMO-based quantum chemical calculations and other fragmentation-oriented methods~\cite{Gordon-2011, Collins-2015, Raghavachari-2015} but also for VQE-UCCSD in general. From the numerical simulations, we also demonstrated that the Trotter error-free UCCSD energy can be estimated by means of extrapolation by computing the UCCSD energies with different numbers of Trotter slices. 

The GPU acceleration was found to be 7.30 and 7.64 with COBYLA and Powell algorithms, respectively, for the dimer ``21'' of (FH)$_3$ (18 qubit system). For the dimer ``21'' of (FH)$_2$-H$_2$O, the estimated GPU acceleration ratio of the VQE quantum circuit simulation to be about 30.85, and the VQE simulations of the dimer ``21'' with COBYLA and Powell will take about 21 and 54 days, respectively.  
The usefulness of GPUs has attracted much attention in various fields of 
quantum computation~\cite{Moller-2017}, and quantum chemistry is an example where the acceleration effect is significant~\cite{Ino-2024}, including in this case; even with GPU acceleration, it still takes orders of magnitude longer than a regular FMO-CCSD(T) calculation~\cite{Mochizuki-2011}.
Recently, an example of large-scale quantum computation with adamantanes has been reported using VQE~\cite{Prasad-2024}. Following these trends, we will perform larger FMO-UCCSD computations on upcoming GPU environments.

\bigskip
\noindent{\Large\bf 5. Acknowledgement} \par
All VQE-UCCSD computations with cuQuantum
on the `Flow' Type-II subsystem at the Information Technology Center of Nagoya University
were performed under the JHPCN Joint Research Projects (jh230001 subject by YM).
KS and YM would like to thank Profs. Takahiro Katagiri (Nagoya University) and Satoshi Ohshima (Kyushu University) for their encouragement. YM was also supported by Rikkyo SFR. 
KS acknowledges the support from Quantum Leap Flagship Program (Grant No. JPMXS0120319794) from MEXT, Japan, Center of Innovations for Sustainable Quantum AI 
(JPMJPF2221) from JST, Japan, and KAKENHI Transformative Research Area B (23H03819) and Scientific Research C (21K03407) from JSPS, Japan.

\newpage

\bibliography{Refs}
\bibliographystyle{unsrt}

\newpage
\noindent{\Large\bf Supplementary Materials} \par

\renewcommand{\thetable}{S1}
\begin{table}[h]
\caption{\label{tab:tableS1} The number of function evaluations in the VQE-UCCSD parameter optimization of (FH)$_3$.}
\begin{center}
\begin{tabular*}{10cm}{@{\extracolsep{\fill}}ccccc}
\hline
 Unit & \multicolumn{2}{c}{CMO} & \multicolumn{2}{c}{LMO} \\
        & COBYLA     & Powell     & COBYLA     & Powell     \\ 
\hline
 Monomer &&&& \\
``1''   & 115  & 225  & 133  & 256  \\ 
``2''   & 130  & 257  & 113  & 257  \\
``3''   & 119  & 193  & 125  & 241  \\
Dimer &&&& \\
``21''  & 1762 & 4783 & 1924 & 5264 \\
``31''  & 1963 & 4765 & 2085 & 5188 \\
``32''  & 1910 & 5408 & 1546 & 5188 \\
\hline
\end{tabular*}
\end{center}
\end{table}

\renewcommand{\thetable}{S2}
\begin{table}[h]
\caption{\label{tab:tableS2} The number of function evaluations in the VQE-UCCSD parameter optimization of (FH)$_2$-H$_2$O.}
\begin{center}
\begin{tabular*}{10cm}{@{\extracolsep{\fill}}ccccc}
\hline
 Unit   & \multicolumn{2}{c}{CMO} & \multicolumn{2}{c}{LMO} \\
        & COBYLA     & Powell     & COBYLA     & Powell     \\ 
\hline
 Monomer &&&& \\
``1''   & 262  & 401  & 474  & 791  \\ 
``2''   & 114  & 238  & 142  & 246  \\
``3''   & 121  & 215  & 129  & 207  \\
Dimer &&&& \\
``21''  & 3278 & 12448 & 4913 & 12452 \\
``31''  & 3511 & 11966 & 4568 & 12764 \\
``32''  & 1348 & 2544 & 1342 & 3882 \\
\hline
\end{tabular*}
\end{center}
\end{table}

\renewcommand{\thetable}{S3}
\begin{table}[h]
\caption{\label{tab:tableS3} The difference of the UCCSD:PW energies from the CAS-CI values of 4H cluster (monomer) and 8H cluster (dimer) calculated with different Trotter decomposition conditions, in units of kcal mol$^{-1}$}
\begin{center}
\begin{tabular*}{16cm}{@{\extracolsep{\fill}}cccccccc}
\hline
Unit   & \multicolumn{5}{c}{First-order Trotter} & Second-order Trotter & \texttt{expm\_multiply}\footnote{Trotter-free implementation using \texttt{expm\_multiply} in SciPy library.} \\
 & $M = 1\footnote{$M$ represents the number of Trotter slices.}$ & $M = 2$ & $M = 3$ & $M = 4$ & $M = 5$ & $M = 1$ & \\
\hline
Monomer     & 0.8102 & 0.8101 & 0.8099 & 0.8100 & 0.8100 & 0.8101 & 0.8118 \\
Dimer (LMO) & 1.6207 & 1.6200 & 1.6203 & 1.6204 & 1.6205 & 1.6206 & 1.6236 \\
Dimer (CMO) & 5.0319 & 1.9139 & 1.7141 & 1.6681 & 1.6496 & 2.0172 & 1.6234 \\
\hline
\end{tabular*}
\end{center}
\end{table}

\renewcommand{\thefigure}{S1}
\begin{figure}[h]
\includegraphics[width=\textwidth]{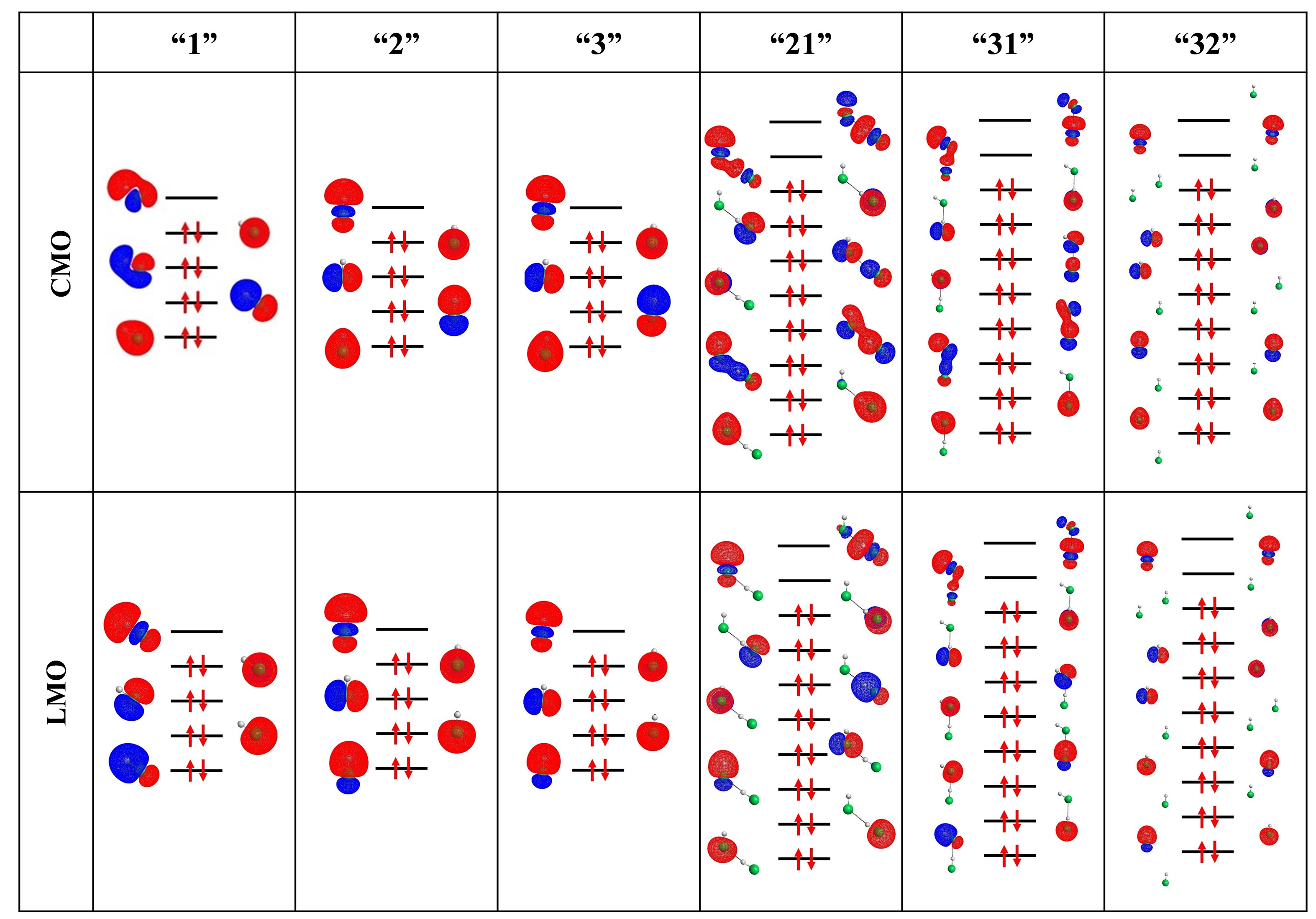}
\caption{\label{fig:figS1} Active orbitals of monomers and dimers of (FH)$_3$. CMO and LMO stand for the HF canonical molecular orbitals and the localized molecular orbitals constructed by using Pipek--Mezey method, respectively. Red arrows specify the electron occupancy of the HF wave function. }
\end{figure}

\renewcommand{\thefigure}{S2}
\begin{figure}[h]
\includegraphics[width=\textwidth]{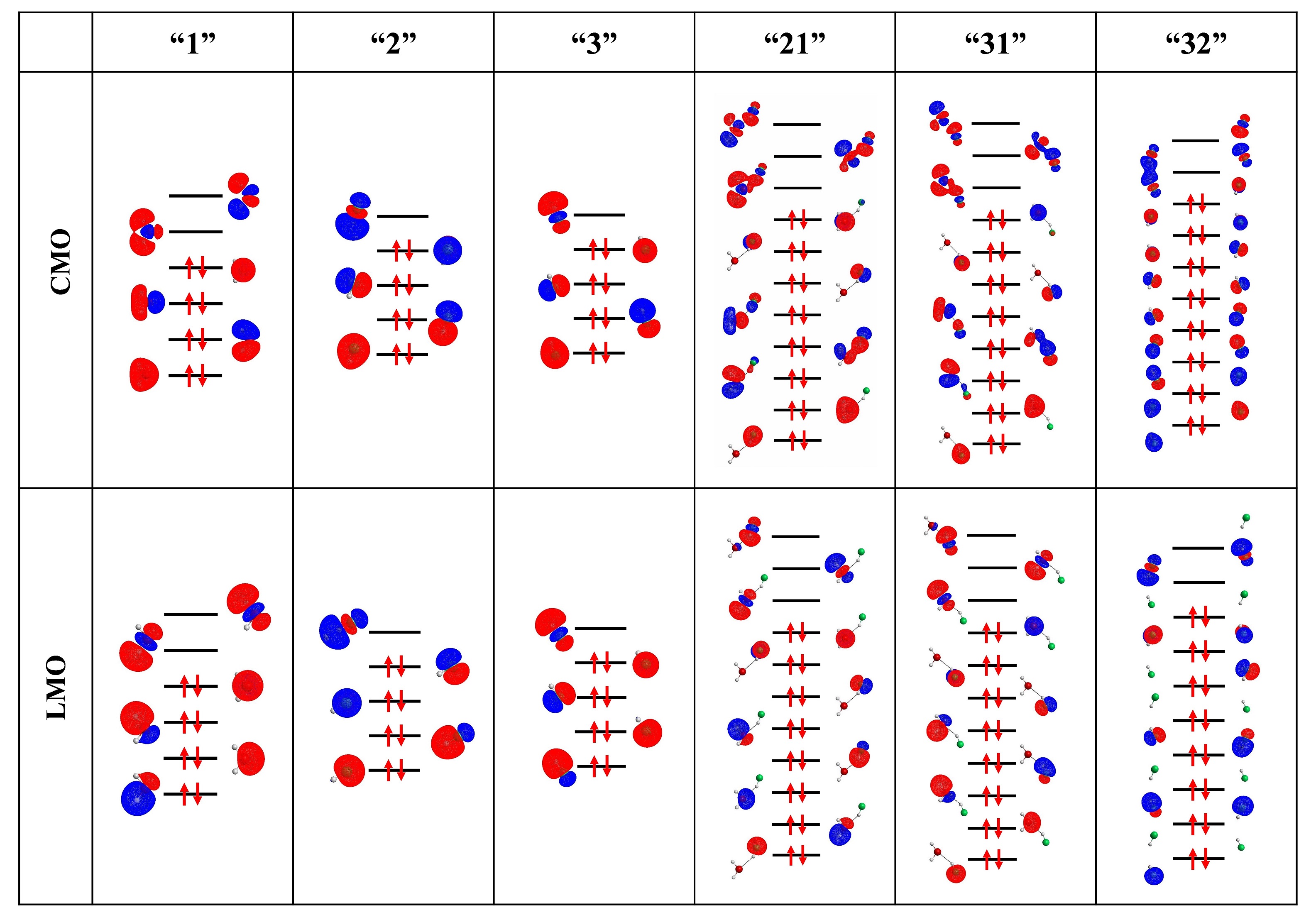}
\caption{\label{fig:figS2} Active orbitals of monomers and dimers of (FH)$_2$-H$_2$O. CMO and LMO stand for the HF canonical molecular orbitals and the localized molecular orbitals constructed by using Pipek--Mezey method, respectively. Red arrows specify the electron occupancy of the HF wave function. }
\end{figure}

\end{document}